Proprioceptive feedback modulates coordinating information

in a system of segmentally-distributed microcircuits

by

Brian Mulloney[1*]

Carmen Smarandache-Wellmann[2]

Cynthia Weller[1]

Wendy M. Hall[1]
and

Ralph A. DiCaprio[3*]

[1] Neurobiology, Physiology, and Behavior
University of California
Davis CA 95616-8519 USA

[2] Emmy Noether Group
Zoological Institute
University of Cologne
D50674 Cologne Germany

[3] Biological Sciences
Ohio University
Athens OH

* Joint senior authors

Running title: Proprioceptive modulation of coordinating neurons

Corresponding Author:

B. Mulloney
NPB
196 Briggs Hall
UC Davis
One Shields Drive
Davis CA 95616-8519 USA
email: bcmulloney@ucdavis.edu
TEL: 530-758-3687

**ABSTRACT**

The system of modular neural circuits that controls crustacean swimmerets drives a metachronal sequence of power-stroke (PS, retraction) and return-stroke (RS, protraction) movements that propels the animal forward efficiently. These neural modules are synchronized by an intersegmental coordinating circuit that imposes characteristic phase differences between these modules. Using a semi-intact preparation that left one swimmeret attached to an otherwise isolated central nervous system (CNS) of the crayfish, *Pacifastacus leniusculus*, we investigated how the rhythmic activity of this system responded to imposed movements. We recorded extracellularly from the PS and RS nerves that innervated the attached limb and from coordinating axons that encode efference copies of the periodic bursts in PS and RS axons. Simultaneously we recorded from homologous nerves in more anterior and posterior segments. Maintained retractions did not affect cycle period, but promptly weakened PS bursts, strengthened RS bursts, and caused corresponding changes in the strength and timing of efference copies in the module's coordinating axons. These changes in strength and timing of these efference copies then caused changes in the phase and duration, but not the strength, of PS bursts in modules controlling neighboring swimmerets. These changes were promptly reversed when the limb was released. Each swimmeret is innervated by two nonspiking stretch receptors (NSSRs) that depolarize when the limb is retracted. Voltage-clamp of an NSSR changed the durations and strengths of bursts in PS and RS axons innervating the same limb, and caused corresponding changes in the efference copies of this motor output.

**INTRODUCTION**

The nervous systems of arthropods and most terrestrial vertebrates contain dedicated neural circuits that control movements of individual limbs distributed in different segments of the CNS. Effective locomotion requires that these circuits be synchronized and coordinated, and that this coordination be sensitive to perturbations of movements of individual limbs. The interplay of central coordinating circuits and proprioceptive feedback to individual limbs during locomotion is dynamic, and the extent to which coordination depends on proprioceptive feedback from each limb in each cycle of movements varies widely between different kinds of animals. This dependence is correlated with the accuracy required of each cycle of movement. Swimming and flying in fluid media, which do not require that the full weight of the body be supported by individual limbs during each cycle, require less accuracy than do, for example, climbing a blade of grass or galloping on rough ground.

The interplay of a central coordinating circuit and proprioceptive feedback can also reveal features of the CNS's operation. The crustacean swimmeret system has a comparatively low requirement for proprioceptive feedback, and fictive swimmeret beating can be recorded in the complete absence of feedback (Hughes and Wiersma 1960; Ikeda and Wiersma 1964). The complex motor pattern that drives each cycle of swimmeret movements involves synchronized output from four pairs of segmental microcircuits, and the intersegmental circuit that coordinates them is now known in cellular detail (Smarandache et al. 2009; Smarandache-Wellmann et al. 2014; Smarandache-Wellmann and Grätsch 2014). This circuit synchronizes oscillations of the microcircuits in different segments and imposes a posterior-to-anterior phase progression that is stable through a wide range of cycle periods. Key elements of this intersegmental circuit are coordinating neurons originating in each microcircuit that encode efference copies of each cycle of its output. How does proprioceptive feedback from individual swimmerets affect the output of this integrated system? At rest, the swimmerets of living prawns and crayfish are rotated anteriorly against the ventral surface of the next anterior abdominal segment (Fig. 1A). During forward swimming, each swimmeret periodically swings posteriorly (retraction) in a power-stroke, and then swings anteriorly (protraction) in a return stroke to its resting position, a periodic cycle of retraction and protraction.

**Sensory innervation of each swimmeret.**

Each swimmeret has its own four sets of sensory neurons (Mulloney and Smarandache-Wellmann 2012). Spiking afferents respond to deflection of setae on the distal rami of each swimmeret (Fig. 1B). Others respond to deformation of the cuticle of these rami (Killian and Page 1992a; b). A third set of spiking afferents and a pair of Nonspiking Stretch Receptors, NSSRs (Heitler 1982), innervate sensory strands that span the joint between the coxa and the basipodite (CB joint) of each swimmeret (Miyan and Neil 1986; Heitler 1986; MacMillan and Deller 1989; Davis 1969). These strands are stretched by retraction or rotation of the swimmeret. The spiking afferents from these strands have peripheral cell bodies and thin axons that project through N1 into the segmental ganglion. The NSSRs, in contrast, have central cell bodies and large-diameter processes that extend from the ganglion's Lateral Neuropil, LN (Skinner 1985b), through the anterior branch of N1 to reach the sensory strand. NSSRs are depolarized by stretching of their sensory strands that occurs during each retraction, and conduct this graded depolarization as an analog signal from their peripheral terminals to their synaptic contacts within the LN (Heitler 1982; 1986).

To study the influence of proprioceptive afferents on the intersegmental coordinating circuit, we mechanically retracted one swimmeret into the position it would reach at the end of each power-stroke, and held it there, while recording from the motor neurons and coordinating interneurons associated with that swimmeret (Fig. 1D). Maintained retraction altered the timing and strengths of bursts of spikes in PS and RS motor axons that innervated the retracted limb. It also altered the timing and numbers of spikes fired in each cycle by the two coordinating neurons that encode efference copies of these PS and RS bursts. These changes in these efference copies affected the durations and phases of bursts of spikes in PS axons that innervated swimmerets in more anterior and more posterior segments, but did not change the strengths of these bursts. The significance of these results and their relevance to earlier work on this system are discussed in the context of the recent description of the intersegmental coordinating circuit.

## MATERIALS AND METHODS

*Isolated nerve cord preparations.* Crayfish were first anesthetized on ice and then exsanguinated by replacing their hemolymph with crayfish saline. Then the ventral nerve cord from the fourth thoracic ganglion, T4, to the last abdominal ganglion, A6 (Fig. 1E) was removed to a Sylgard-lined dish filled with normal saline. Normal saline contained (in mM) 5.4 KCl, 2.6 $MgCl_2$, 13.5 $CaCl_2$, 195 NaCl, buffered with 10 mM Tris base and 4.7 mM maleic acid at pH 7.4. The sheaths were removed from the dorsal side of each ganglion. Isolated ventral nerve cord preparations sometimes spontaneously express the normal swimmeret motor pattern (Hughes and Wiersma 1960), but in other preparations that were either intermittently active or were silent, a 1.5 – 3 μM solution of the cholinergic agonist carbachol (Sigma) in normal saline was perfused over the preparation to elicit continuous expression of a stable motor output (Mulloney 1997; Braun and Mulloney 1993; Chrachri and Neil 1993).

*Recording methods.* To record action potentials in the axons of power-stroke (PS) and return-stroke (RS) motor neurons, pin electrodes were placed on the anterior and posterior branches of the swimmeret nerves, N1, that project from each ganglion A2 through A5 (Fig. 1E) (Mulloney and Hall 2000). To record activity of the intersegmental coordinating neurons, $ASC_E$ and DSC, that originate in each module (Smarandache et al. 2009), we placed a suction electrode on the Minuscule Tract (MnT) as it crossed dorsal to the Lateral Giant axon (Mulloney et al. 2003; Skinner 1985a; Smarandache et al. 2009). All extracellular recordings were band-pass filtered and amplified with A-M Systems 1700 high-gain preamplifiers (Carlsborg WA).

Intracellular recordings were made with npi SEC-05 amplifiers (npi electronic, Tamm, Germany) from processes of neurons within the Lateral Neuropil (LN) using sharp glass microelectrodes. During each microelectrode experiment, neurons were tentatively identified by their physiological properties, and then filled with an intracellular marker. This tentative identification was later confirmed or contradicted by the neuron's anatomy. Microelectrodes contained a solution of 1 M $K^+$ acetate + 0.1 M KCl and 1% Dextran Texas Red, dTR (Dextran Texas Red MW 3000 lysine fixable; Life Technologies, Grand Island NY). These electrodes had tip resistances of 30-50 Megohms. Neurons were filled iontophoretically with +1.0 nA current steps 0.25 sec long at 2 Hz for 20 minutes.

Extracellular and intracellular recordings were digitized, usually at 10 kHz, using an Axon Instruments Digidata 1322A and pClamp software (Molecular Devices, Union City CA), and saved as computer files for later analysis.

*Semi-intact preparations.* To study the influence of a swimmeret's movements on the motor output to it and to other swimmerets, we developed a semi-intact preparation in which one swimmeret remained attached to the CNS (Fig. 1D). The ventral nerve cord (T4 through A6) was removed and pinned out dorsal side up as described above, except the lateral abdominal wall and swimmeret of one segment, either segment 3 or 4, were left attached through N1 to ganglion A3 or A4. To allow the swimmeret to move, the piece of body wall was rolled 180° around the N1 and pinned securely to the Sylgard. The swimmeret itself was shortened by cutting off its distal rami so their movements would not hit the electrodes. The major PS and RS muscles of the swimmeret originate on the medial surface of the lateral abdominal wall (Fig. 1B), and in some of these preparations the swimmeret beat periodically in what appeared to be mechanically normal cycles of movements. Pin electrodes were placed on the anterior and posterior branches of the twisted N1 to record PS and RS spikes *en passant* (Fig. 1D). A suction electrode was placed on the MnT above the module innervating the attached swimmeret to record firing of coordinating neurons that originated in that module. In some experiments, a microelectrode was then brought in to record intracellularly from neurons in the module that innervated the attached swimmeret.

The outer end of the basipodite of the swimmeret itself was grasped by a hook fashioned from an insect pin attached to a micromanipulator or to the headstage of a puller controlled by a proportional-integro-differential controller (Aurora Scientific Model 322C, Aurora, Ontario, Canada) operating in length-feedback mode. Commands to move the swimmeret using the controller were generated with a sine-wave generator or with waveforms generated by pClamp (Molecular Devices, Union City CA).

*Measuring parameters of the motor patterns.* The start and stop times of bursts of spikes in extracellular recordings were measured using Dataview (http://www.st-andrews.ac.uk/~wjh/) and used to describe the temporal structure of the periodic motor pattern. The period of each cycle was the interval from the start of one PS burst to the start of the next PS burst. The latency of an event occurring during a cycle was measured as the time interval between the start of the event and the start of the preceding PS burst that marked the start of the cycle. The phases of these events were then defined as the ratio of these latencies to that cycle's period. Therefore, phase could range from 0 to 1.0.

*Measuring strengths of bursts of spikes.* In different cycles of motor output, the strengths of bursts of spikes in PS and RS motor neurons – the numbers of motor neurons recruited and the numbers of spikes each neuron fired (Davis 1971) – could vary. These strengths were measured using a modification of the method detailed in Mulloney (2005). First, the mean voltage was subtracted from each trace to remove any DC offset and the voltage at each time-step was squared to rectify it. The rectified trace was then smoothed using a Fast Fourier Transform (FFT) with a triangular kernel. The half-width of this kernel was 327 time-points, corresponding to 32.7 msec sampled at 10 kHz. The smoothed recording was restored using an Inverse FFT. Then, using the lists of times at which each burst started and stopped, which were measured independently using Dataview, we isolated the smoothed waveform of each burst and calculated its area. Dividing each burst's area by its measured duration gave a measure of its strength that increased as the numbers and sizes of spikes within it increased (Mulloney 2005). This procedure was effective even when an electrode recorded bursts of spikes in more than one functional group of motor axons, e.g. the RS4 trace in Figure 1F.

*Statistical analysis.* To test the probability that a parameter measured under two conditions during an individual experiment did not change, we calculated t-tests (Zar, 1996) using the routines in SigmaPlot 12.5 (SysStat, San Jose CA). To test the probability that a parameter measured under two conditions did not change in a group of similar experiments, we calculated Paired t-tests (Zar, 1996) using SigmaPlot. Results of these calculations are reported as Probabilities plus the Power of each calculation.

*Confocal microscopy and imaging procedures.* At the end of each experiment, the preparation was fixed overnight in 4% paraformaldehyde in phosphate buffered saline (PBS). Preparations were then rinsed with PBS four times for 10 min each, pinned out dorsal-side up, and cleared in an ascending ethanol series to methyl salicylate. Cleared whole mounts were mounted in methyl salicylate in a Permanox® dish with a cover-slip base. To prevent movement during imaging, a small, thick slip of glass was placed on top of the nerve cord.

Preparations were examined as whole mounts oriented for frontal view, dorsal side up (Fig. 1E). The structure of each labeled neuron was captured as a stack of confocal images that extended from the most dorsal to the most ventral part of the cell. Images were captured using an Olympus FLUOVIEW 300 confocal microscope (Olympus America Inc, Center Valley, PA) equipped with krypton (488 nm) and argon (568 nm) lasers, and an Olympus 20x 0.7 NA UPlanApo lens. Step size was 0.75 µm. The images were converted to 24-bit TIF images in Fluoview software, where the gamma and intensity were adjusted to optimize the background intensity. The resulting images were then transferred to Adobe Photoshop for further adjustment of brightness, contrast, and sharpness. All images in each stack were adjusted uniformly in the same way.

## RESULTS

### Movements imposed on a swimmeret affect the motor output to that swimmeret.

In semi-intact preparations with a swimmeret attached (Fig. 1D), we recorded extracellularly from coordinating axons and power-stroke (PS) and return-stroke (RS) motor axons while moving the swimmeret. In nine preparations in which the system was actively expressing continuous, normally-coordinated motor output, moving the limb from its resting protracted position to a fully-retracted position immediately changed the relative strengths of bursts of spikes in the PS and RS branches of the N1 that innervated the retracted swimmeret (Fig. 2A). PS bursts became shorter and weaker while RS bursts became longer and stronger. When this retraction was halted and the limb returned to its fully protracted resting position (Fig. 2B), PS bursts and RS bursts promptly returned to same levels they had before the retraction was imposed. Both in isolated CNS preparations and in these semi-intact preparations, RS activity was commonly quite weak. To quantify changes caused by retraction of a swimmeret, we selected from the nine experiments two in which the RS activity was strong, and compared parameters of the PS and RS bursts produced when the limb was in its resting position and when it was fully retracted, for three retractions in each experiment. During every retraction ($n = 6$ expts.), durations of PS bursts were shorter (One-tailed t-test $P < 0.001$, power $\geq 0.996$), and RS bursts were longer (One-tailed t-test $P < 0.001$, power $= 1.0$) during the retraction than they were before the retraction began. These same PS bursts were also weaker (One-tailed t-test $P \leq 0.014$, power $\geq 0.726$) and the same RS bursts were stronger (One-tailed t-test $P \leq 0.013$, power $\geq 0.623$) during each retraction than those recorded before the retraction began.

These retractions did not affect the period of the system's motor output. During retraction, the mean periods were 1 msec shorter than when the limb was in its resting position (Paired t-test P = 0.437, n = 9 expts, power = 0.052). In other experiments where the swimmeret was moved sinusoidally at periods close to the period of the preparation's expressed motor output, we did not observe entrainment of that output.

**Imposed retractions also affected firing of coordinating neurons.**

$ASC_E$ neurons in each module encode information about the timing, duration, and strength of each burst in the module's population of PS neurons. DSC neurons encode similar information about each burst in the population of RS neurons (Mulloney et al. 2006). During retractions of a swimmeret, bursts of spikes in the coordinating axons that originated in the same module also changed (Figs 2A, 2B). The numbers of spikes per $ASC_E$ burst decreased along with the decreased strengths of PS bursts (Fig. 3A). At the same time, numbers of spikes per DSC burst increased as the strengths of PS bursts decreased (Fig. 3B). In five experiments, $ASC_E$ burst durations were shorter during retraction while DSC burst durations were longer (Table 1), and the phases of both $ASC_E$, and DSC bursts relative to PS4 also advanced (Fig. 4). These correlated responses of PS motor neurons and the $ASC_E$, and DSC coordinating neurons suggest that the proprioceptive afferents that respond to retraction and protraction of a swimmeret affect the module's pattern-generating kernel, not just the motor neurons themselves.

**Retractions of one swimmeret also affected motor output to neighboring swimmerets.**

Bursts of spikes in $ASC_E$ and DSC coordinating neurons synchronize neighboring swimmeret modules (Zhang et al. 2014; Smarandache-Wellmann et al. 2014). Changes in the strengths or timing of these bursts affect the strengths and phases of the motor output from modules in neighboring segments (Mulloney and Hall 2007). Given this background, we compared the phases, durations, and strengths of PS bursts from more anterior and posterior segments when one swimmeret was held in its resting protracted position and when it was held in its fully retracted position and thereby altered firing of its own module's coordinating neurons.

The consequences of a retraction for the PS motor neurons innervating the retracted swimmeret were different from the consequences for their homologues in neighboring segments (Figs. 1E, 4). Both the durations and the strengths of bursts in PS motor neurons innervating the retracted limb decreased significantly during retractions (Fig. 2; PS4 in Table 1). Simultaneously, durations of PS bursts in more anterior segments also became shorter (PS3 in Table 1), while durations of PS bursts in more posterior segments became longer (PS5 in Table 1). In addition to the phases of DSC and $ASC_E$ (Fig 4), the phases of both anterior and posterior PS bursts were advanced during a retraction. But, in contrast to the strengths of bursts in the PS motor neurons innervating the retracted limb (Fig. 2D), the strengths of PS bursts in these other segments were unchanged (Table 1). Therefore, a retraction's major effect on neighboring modules was on the timing, not the strength, of their PS motor output.

**Nonspiking stretch receptors (NSSRs).**

These results raise questions about proprioceptive feedback from a swimmeret to the module that innervates it. In two series of ablation experiments, Davis (1969) and MacMillan and Deller (1989) demonstrated that the sensory strands innervated by NSSRs are necessary and sufficient to elicit reflex responses to imposed swimmeret movements. We therefore focused our attention on these NSSR neurons (Fig. 5).

When the system is actively expressing the normal swimmeret motor pattern, the membrane potentials of these receptors oscillate periodically in phase with the motor output because they receive synaptic input from some components of the module's pattern-generating kernel (Paul 1989). When a swimmeret was retracted in semi-intact preparations that were actively expressing normal motor patterns, recordings from the central processes of an NSSR innervating the retracted swimmeret showed a proportional depolarization superimposed on the periodic oscillations caused by periodic synaptic currents (Fig. 6). Retraction of the NSSR through a wider angle caused a larger depolarization and greater inhibition of the PS motor output (Heitler 1982).

**Depolarization of an individual NSSR altered its module's motor output.**

To explore the contributions of individual NSSR neurons to proprioceptive modulation of a swimmeret module's output, we made microelectrode recordings from NSSRs in the LN of isolated CNS preparations (Figs. 1E, 5) that were actively producing coordinated swimmeret motor patterns. Step depolarizations of an NSSR with current injections caused immediate weakening of PS bursts that promptly recovered when the step ended (Fig.7A). Step changes in potential using dSEVC affected both PS and RS activity in the same module (Fig. 7B). Hyperpolarization to -90 mV, about 30 mV below resting potential, increased durations of PS bursts but shortened and weakened RS bursts (Fig. 7C). Depolarization to -30 mV, about 30 mV above rest, shortened and weakened PS bursts, but strengthened and lengthened RS bursts (Fig. 7C). During these depolarizations, new larger RS units were recruited, as has been described in response to increases in excitation of the system (Davis 1971; Mulloney 1997).

Depolarizing an NSSR also affected coordinating neurons in the same module (Table 2). $ASC_E$ bursts lost one spike per cycle, but their durations did not change significantly. DSC bursts, however, gained 4 spikes per cycle and lasted 80 msec longer (Table 2). The phases of both $ASC_E$ and DSC bursts did not change during depolarization of the NSSR.

Each swimmeret module has two NSSRs, NSSR-A and NSSR-P (Heitler 1982), that differ in the positions of their cell bodies. In all of our experiments, we filled the recorded neuron and identified it as one of these types. Tabulating the experiments by NSSR type failed to reveal any physiological difference between them, and we conclude that they are functional equivalents operating in parallel.

**How does this work?**

Both depolarization of an NSSR and retraction of a swimmeret appeared to affect all the PS and RS motor neurons and both coordinating neurons within the same microcircuit similarly, so we think it is likely that these effects are caused by synaptic connections between the NSSRs and components of the microcircuit's pattern-generating kernel (Smarandache-Wellmann et al. 2013; Smarandache-Wellmann et al. 2014).

One well-established pathway through which coordinating information from other modules tunes the strength and phase of a microcircuit's output is the connection between ComInt 1 and the IRSh neuron in the microcircuit's pattern-generating kernel (Smarandache-Wellmann et al. 2014), but this is not the pathway through which NSSRs operate. Dual microelectrode recordings from ComInt 1 neurons and NSSRs in the same module revealed no evidence of connections between these neurons. Depolarizing an NSSR strongly enough to inhibit PS motor bursts completely did not affect ComInt 1's membrane potential (data not shown, N = 3 expts). Depolarizing or hyperpolarizing ComInt 1 strongly enough to affect the microcircuit's PS and RS motor output reduced the amplitude of NSSR's oscillations, but there was no evidence of time-locked post-synaptic responses (data not shown).

The kernel of each swimmeret module is composed of two classes of nonspiking local interneurons, IPS and IRS, which respectively inhibit PS and RS motor neurons (Smarandache-Wellmann et al. 2013; Mulloney 2003). IPS and IRS neurons are connected by reciprocal inhibition to form the module's pattern-generating circuit. During imposed retractions of a swimmeret, the membrane potential of one type of IPS neuron, IPSt, depolarized as the limb was retracted (Fig. 8). IPSt is a component of the microcircuit's pattern-generating kernel, so this observation is consistent with the idea that NSSRs synapse with the module's pattern-generating neurons.

## DISCUSSION

Maintained retraction of a swimmeret causes a differential alteration of the periodic motor output that would drive its unrestricted movements. These alterations would increase the strengths of contractions of return-stroke muscles that oppose the retraction and decrease the strengths of contractions of power-stroke muscles that assist the retraction. The two coordinating neurons, $ASC_E$ and DSC, that originate in the microcircuit innervating the retracted limb are also differentially affected (Table 1). These two coordinating neurons change the phases and numbers of spikes in each of their bursts to track changes in the timing and strengths of the different motor bursts encoded by each neuron. Maintained retraction affects the timing of PS output both to more anterior and to more posterior swimmerets, consistent with the altered firing of these $ASC_E$ and DSC neurons (Table 1).

**Entrainment of the swimmeret motor pattern by imposed movements.**

It is remarkable that the periodic alternation of PS and RS bursts to the retracted swimmeret did not halt during these maintained retractions (Figs. 2, 3); indeed, the period of the pattern did not change at all. This robust resistance to entrainment by movements of one swimmeret has been consistently reported in earlier studies (Heitler 1986), but see MacMillan and Deller (1989). In pioneering experiments that used command neurons (Acevedo et al. 1994; Wiersma and Ikeda 1964) to elicit swimmeret beating from preparations which had all swimmerets attached, West et al. (1979) found that mechanical interference with one swimmeret sometimes decreased the cycle period, but these effects were variable. In hindsight this variability could be attributed both to incomplete retractions using free-hand interference and to variability in the method used to elicit swimmeret beating. Only by moving several swimmerets simultaneously and synchronously were Deller and MacMillan (1989) able to entrain the system's output. Given that the system consists of four pairs of microcircuits, each with its own pattern-generating kernel, linked together by a coordinating circuit that distributes weighted information about each microcircuit's status to all the others, the robustness of period and relatively small changes in phase is understandable (Smarandache et al. 2009; Smarandache-Wellmann et al. 2014).

**Each swimmeret's pair of NSSRs contributes to these responses to retraction.**

Full retraction of a swimmeret should stretch all of the limb's position-sensitive afferents, and probably also stimulate some of its cuticular stress receptors. The sensory transduction apparatus of both NSSRs is inserted into an elastic strand that spans the swimmeret's CB joint and is stretched by retraction. Retraction of a swimmeret depolarizes the NSSRs that signal its position (Heitler 1982; 1983; 1986; MacMillan and Deller 1989; Miyan and Neil 1986). Although no detailed analysis of the transfer properties of swimmeret NSSRs is available, previous work indicates that the graded depolarizations of these afferents (Fig. 6) are relatively linearly proportional to receptor length, at least for 0.5-6 Hz sinusoidal movements (Heitler 1982). Rapid stretches produce graded depolarization that again tracks receptor length, plus an additional larger transient depolarization (Heitler 1982). These transient receptor potentials might underlie the transient responses to retraction that we see both in PS burst strengths and in numbers of spikes per burst in $ASC_E$ and DSC neurons (Fig. 3).

We found that voltage-clamped depolarization of individual NSSRs simultaneously inhibited PS firing somewhat and excited RS firing (Fig. 7C). Earlier ablation experiments indicated that the spiking afferents from setae on the distal rami of a swimmeret (Fig. 1B) played at most minor roles in modulating the force of swimmeret movements (Davis 1969; MacMillan and Deller 1989), but that sinusoidal currents injected into an NSSR could modulate firing of swimmeret motor neurons in the same microcircuit (Heitler 1986). So, the physiology of individual NSSRs can account for some, but not all, features of a microcircuit's responses to retraction. Recall that each microcircuit has two NSSRs. Retraction should depolarize both NSSRs simultaneously, allowing them to act in parallel on their postsynaptic targets. By using a Vaseline-gap stimulation method that would depolarize both NSSRs simultaneously, MacMillan and Deller (1989) were able to entrain the system's output to the period of imposed voltage oscillations.

**Comparisons with proprioceptive innervation of other decapod limbs.**

Like the movements of each swimmeret, movements of each walking leg and each gill bailer of decapod crustacea are monitored by proprioceptors at their bases that include both spiking and nonspiking receptor neurons. The thoracic coxal muscle receptor organ (TCMRO) spans the thoracic-coxal joint and has two non-spiking afferents (Bush and Roberts 1971) while the levator and depressor receptors span the coxal-basal joint and are innervated by two and one non-spiking afferents respectively (Cannone and Nijland 1989; Cannone 1987), along with the spiking coxal-basal chordotonal organ (CBCTO). Each gill-bailer, "scaphognathite", also has a singe intrinsic proprioceptor, the oval organ (OO), that has three afferent neurons(Bush and Pasztor 1983; Pasztor and Bush 1983).

The transfer functions of these different nonspiking afferents in the TCMRO and the CB receptor strands are nonlinear (DiCaprio 2003), although the relationship between receptor potential and receptor length are relatively linear for 0.5-5Hz sinusoidal length changes (DiCaprio, unpublished data) . Like the swimmeret NSSRs, the nonspiking afferents in the OO receive a strong synaptic modulation, in phase with the ventilatory motor pattern, from the local pattern-generating circuit (DiCaprio 1999). This modulation is inhibitory, and gates sensory input to the local circuit in a phase dependent manner.

All the spiking and nonspiking receptors at the TC and CB joints of crab walking legs are broad-band receptors capable of signaling at frequencies up to 200 Hz. However, the information transfer rate of the nonspiking afferents is about ten times greater than the maximum rate of the spiking CBCTO afferents over the same bandwidth. These high information transfer rates result in very high signal-to-noise ratio (SNR) of the nonspiking afferents at stimulus frequencies up to and above 200 Hz (DiCaprio 2004; DiCaprio et al. 2007). Low-frequency stimulation of nonspiking CBCTO receptors produces changes in membrane potential with SNRs as great as those achieved by these neurons at high frequencies, unlike the CBCTO's spiking afferents. By comparison, the depolarization of a swimmeret's NSSRs would probably produce a faithful representation of the swimmeret's movements throughout its normal range of frequencies, less than 10 Hz.

**Comparisons with proprioceptive modulation during other forms of locomotion.**

Because large-tailed crustaceans swim in a continuous fluid medium, and all but the largest are close to neutrally-buoyant, forward swimming using the swimmeret system does not face some of the same mechanical problems that influence walking in terrestrial environments. During walking, both the load borne by each leg and the leg's position relative to the body are monitored by proprioceptors. These proprioceptors have major influences on each step, and on the coordination of these steps with those of other legs. In insects walking in different directions through an environment with different barriers, sensory feedback plays a crucial role in coordinating leg movements (Büschges 2008). In stick insects, independent of the walking direction, the motor program switches from swing-phase to stance-phase when the leg's load sensors (campaniform sensilla) or position sensors (femoral chordotonal organ) are stimulated. When both sensory channels are activated simultaneously, a faster, more prolonged inhibition of the leg's flexor motor neurons is elicited (Akay and Büschges 2006). The role of NSSRs in the swimmeret system seems more like that of position sensors of the hip joint in walking mammals, which have major influences on each cycle of stepping. Prolonged retraction of a walking cat's hind leg alters the relative strengths of flexor and extensor motor output, and also affects the timing of coordinated motor output to the other legs (Rossignol et al. 2006).

In the normal course of periodic swimmeret movements, the NSSRs of each swimmeret are stretched only late in each power-stroke, and the extent of each stretch is determined by the angle of rotation reached at the end of the power-stroke. Therefore, the NSSRs might add well-timed, proportional excitation that augments each RSE burst and the subsequent return-stroke, much as stretch receptors of locust wings contribute pulses of excitation that shape the centrally-driven bursts of spikes in wing depressor motor neurons (Burrows 1975; Pearson and Wolf 1988) and influence the periods of wing beats during flight (Wilson and Gettrup 1963; Möhl 1985; Büschges and Pearson 1991).

In summary, imposed retraction of one swimmeret has both local and distant effects on the ongoing activity of the swimmeret system. The local effects include weakening the drive to PS muscles and strengthening the drive to RS muscles as long as the retraction continues, both of which oppose the ongoing retraction. The local effects also change the timing and strengths of bursts of spikes in the local circuit's $ASC_E$ and DSC neurons, which encode three parameters of each cycle: when each PS and RS burst began, how long it lasted, and how strong it was (Mulloney et al. 2006). We think these quantitative changes in $ASC_E$ and DSC bursts, changes in the efference copies of each cycle, are decoded by their postsynaptic targets in other segments and cause their temporal responses to movement of one limb (Smarandache-Wellmann et al. 2014; Smarandache et al. 2009).

**OTHER ACKNOWLEDGEMENTS**

**GRANTS**

This research has been supported by NSF grants 0905063 and NSF 1147058 to BM, and by NIH Grant NS048068 to BM, by an Emmy Noether DFG grant SM 206/3-2 and a startup grant of the University of Cologne for female faculty to C S-W, and by

The confocal microscope used in this study was supported by NIH P30 EY12576.

**DISCLOSURES**

The authors declare they have no known conflicts of interest that influence or could be perceived to influence their work.

**ROLES OF AUTHORS**

BM, CSW, and RAD participated in experiments, data analysis, and writing the paper. WMH participated in experiments and data analysis. CW participated in analysis, preparation of figures, and writing the paper.

Table 1.

Differences in Burst Parameters during Retraction and Protraction of a Swimmeret†

Differences (Retract – Protract) of Means ± SD (number of experiments)

P = One-tailed Probability from Paired t-test

| Neurons | Duration (msec) | Strength | Phase (relative to PS4) |
|---|---|---|---|
| PS3 | -18.5 ± 10.7 (8)<br>$P < 0.001$ | 0.077 ± o.142 (8)<br>$P = 0.083$ | -0.049 ± 0.030 (7)<br>$P = 0.002$ |
| $ASC_E4$ | -72.3 ± 23.7 (5)<br>$P = 0.001$ | nd | -0.029 ± 0.009 (5)<br>$P = 0.001$ |
| PS4 | -64.6 ± 32.5 (8)<br>$P < 0.001$ | -0.996 ± 0.901 (7)<br>$P = 0.008$ | 0 |
| DSC4 | 84.2 ± 37.5 (5)<br>$P = 0.004$ | nd | -0.103 ± 0.043 (5)<br>$P = 0.003$ |
| PS5 | 21.1 ± 12.6 (5)<br>$P = 0.010$ | -0.024 ± 0.271 (5)<br>$P = 0.425$ | -0.062 ± 0.028 (4)<br>$P = 0.011$ |

† Data from experiments in which a swimmeret innervated by ganglion A4 was retracted.

nd: not calculated

Table 2.

Differences in Burst Parameters Caused by Depolarizing an NSSR neuron.[†]

Differences of Means (Depolarized – Rest) ± SD

P = One-tailed Probability (Depol ≥ Rest) from Paired t-test,

except t-test for unpaired DSC data

| Neurons in A4 | Duration (msec) | Spikes per Burst | Phase (relative to PS5) |
|---|---|---|---|
| $ASC_E$ (n = 6) | 1.3 ± 12.3<br>P = 0.404 | -0.735 ± 0.933<br>P = 0.056 | 0.0 |
| PS4 (n = 7) | -42.8 ± 29.9<br>P = 0.009 | nd | 0.029 ± 0.032<br>P = 0.028 |
| DSC (n = 1) | 79.9<br>P < 0.001 | 4.4<br>P < 0.001 | 0.042<br>P= 0.200 |

[†] Data from neurons in ganglion A4 in experiments where an A4 NSSR neuron was depolarized by current injection.

nd: not determined

**FIGURE LEGENDS**

Fig. 1. *A:* A lateral view of a prawn that shows the pairs of swimmerets distributed on segments of the abdomen. From Huxley (1880). *B.* Drawing of one pair of swimmerets that shows their attachments to the abdominal wall, intrinsic and extrinsic musculature, and the innervation of each swimmeret by the nerve (N1) that projects from the segmental ganglion (A3). From Keim (1915). *C.* A drawing of the ventral side of a female crayfish abdomen that shows the four pairs of swimmerets in their resting, protracted positions. One swimmeret, on the right side of abdominal segment 4, has a hook attached to retract it posteriorly through the arc it makes during each cycle of power-stroke and return-stroke movements (double-headed arrow). *D.* Drawing of the semi-intact preparation with one swimmeret still attached that shows the positions of the pin electrodes on the power-stroke (PS) and return-stroke (RS) branches of the N1 innervating the attached swimmeret. To permit swimmeret movements (double-headed arrow) and simultaneous recordings with a microelectrode (ic) in the Lateral Neuropil and a suction electrode (ec) on the Minuscule Tract, the N1 is rolled and the distal rami have been pruned off. *E.* Diagram of the crayfish CNS showing the Brain (B), the thoracic ganglia (T), and the chain of abdominal ganglia (A1 through A6). The abdominal chain is expanded to show the segmental nerves that innervate swimmerets. *F.* Simultaneous recordings from power-stroke (PS) and return-stroke (RS) branches of one N1 in ganglion A4, and from axons of ascending (ASC4) and descending (DSC4) coordinating neurons arising from the same hemiganglion.

Fig. 2. Changes in the motor output from a swimmeret microcircuit caused by mechanical retraction of its swimmeret. *A.* Simultaneous recordings from PS and RS branches of the N1 that projects to the swimmeret, and from the DSC coordinating axon that projects posteriorly from the module. The swimmeret was free to move for the first four cycles and then was held

retracted (horizontal bar). The additional activity in the RS recording that began immediately upon retraction includes sensory afferents that had been silent. *B.* In a different preparation, simultaneous recordings from PS and RS branches of the N1 that projects to the swimmeret, and from ASC axons that project anteriorly from the module. The swimmeret was held retracted during the first five cycles of activity (horizontal bar), and then released. The smaller spikes that begin each ASC burst are from the $ASC_E$ axon, while the larger spikes in each ASC burst during retraction are from the $ASC_L$ axon (Mulloney and Smarandache-Wellmann 2012). *C.* Upon maintained retraction, durations of RS bursts increased while durations of PS bursts decreased. These differential changes in durations stopped once the swimmeret was released. *D.* Retraction of the swimmeret also affected the strengths of PS and RS bursts. PS bursts were weaker and RS bursts were stronger during each retraction. The normal strengths of both PS and RS were restored once the retraction ended. In C and D, the data have been normalized to the Means of the bursts preceding the retraction, the thin horizontal line marks no change, and the dark grey lines are lowess fits to the experimental data.

Fig. 3. Differential responses of $ASC_E$ and DSC axons to three imposed retractions (retract) of a swimmeret during continuous series of PS bursts (Figs. 1F, 2). *A.* During each retraction, PS bursts were weaker, and the number of $ASC_E$ spikes per burst decreased. The PS bursts that immediately followed the end of each retraction were transiently stronger than average. *B.* During each retraction, PS bursts were weaker and the numbers of DSC spikes per burst increased. DSC bursts immediately following the start of each retraction were transiently stronger than average.

Fig. 4. Box plots that compare phases and duty-cycles of PS bursts in ganglia A3, A4, and A5 (PS3r, PS4r, and PS5r), and of $ASC_E$4r and DSC4r bursts recorded simultaneously on the right

side of segment A4 during maintained protraction and retraction of a swimmeret. Blue boxes illustrate the mean duty-cycle of bursts with the limb protracted, in its rest position. Yellow boxes illustrate the mean duty-cycle with the limb held fully retracted. Phases are defined within the cycle of PS4 bursts. Each box begins at the mean phase. Left-hand error bars show Standard Deviation (SD) of phase; right-hand error bars show SD of duty cycle.

Fig. 5. Whole mount of an abdominal ganglion, viewed from the dorsal side with anterior at the top, in which two NSSR neurons were filled with fluorescent dye. On the left side, an NSSR with a posterior cell body (NSSR-P) has small branches from its main process restricted to the left LN, and its axon projects into the left swimmeret nerve, N1. On the right side, an NSSR with an anterior cell body (NSSR-A) has small branches restricted to the right LN, and its axon projects into the right swimmeret nerve.

Fig 6. Intracellular recording from an NSSR, shown as a whole mount at the right, within the LN during expression of normal swimmeret motor output and a slow retraction-protraction of the swimmeret through its full range of movement. The NSSR's membrane potential oscillates periodically in synchrony with the motor output (PS). These oscillations are superimposed on a larger depolarization caused by retraction of the swimmeret. The orientation of the whole mount is the same as that in Figure 5.

Fig. 7. Responses of a module's motor output to changes in the membrane potential of one NSSR. *A.* Depolarization of one NSSR with injected current partially inhibits periodic bursts in PS axons. *B.* Simultaneous recordings of PS and RS activity during voltage clamp of an NSSR that innervated the same swimmeret. The neuron was clamped at -94 mV *(Bi)*, 33 mV more hyperpolarized than its resting potential, -61 mV (Bii), and at -29 mV (Biii), 33 mV more depolarized. *C.* Plots of the durations and strengths of PS and RS bursts recorded in two

experiments (black, grey) in which an NSSR was voltage-clamped at three different potentials. The durations and strengths of RS bursts increased while those of PS bursts decreased as NSSR's membrane potential decreased.

Fig. 8. Simultaneous recordings of the membrane potential of an IPSt neuron, shown as a whole mount at the right, and of PS activity in the same module during a cycle of retraction of the swimmeret. As the swimmeret was retracted, IPSt was depolarized and the PS motor neurons were increasingly inhibited. The orientation of the whole mount is the same as that in Figure 5.

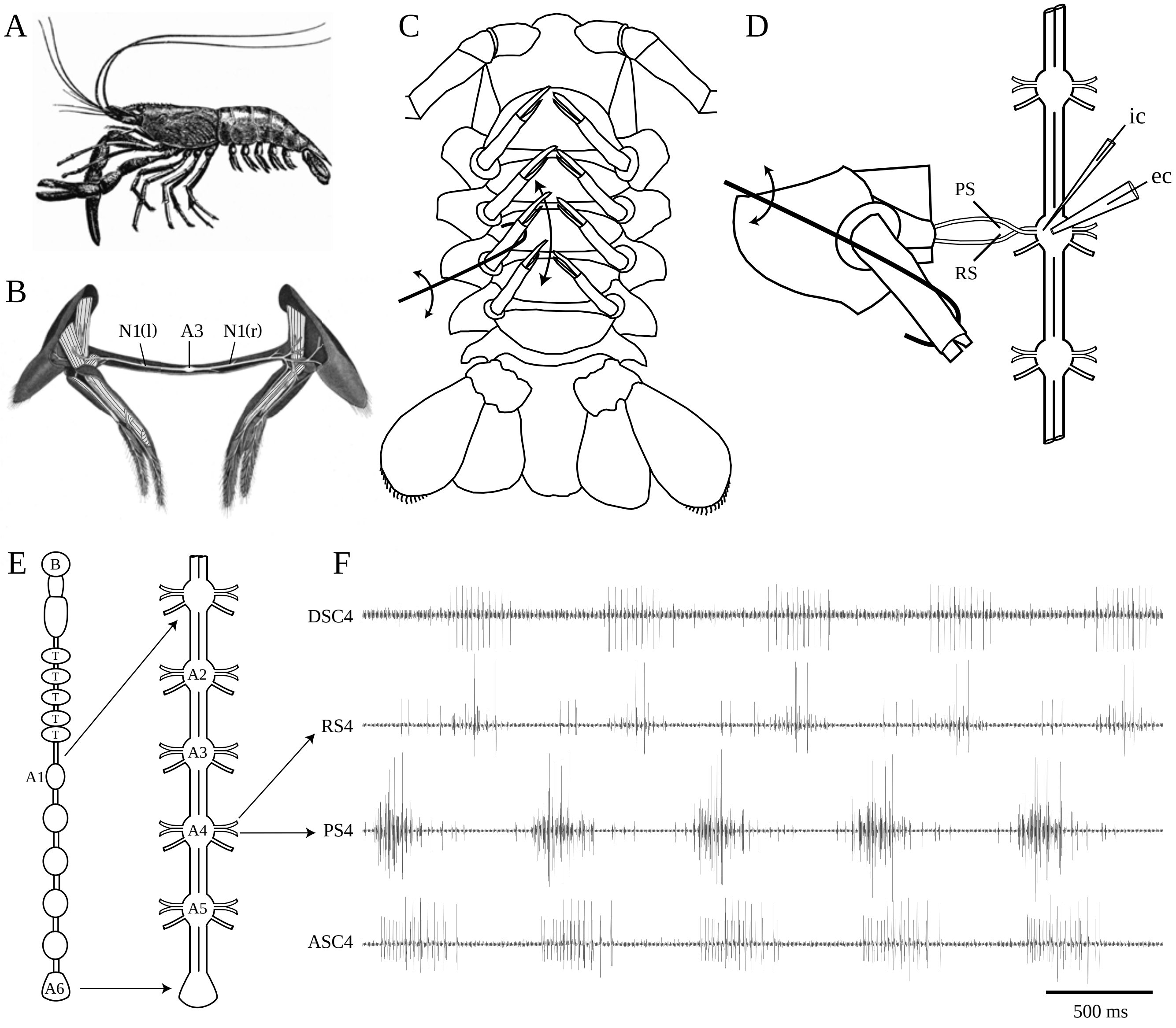

Figure 1

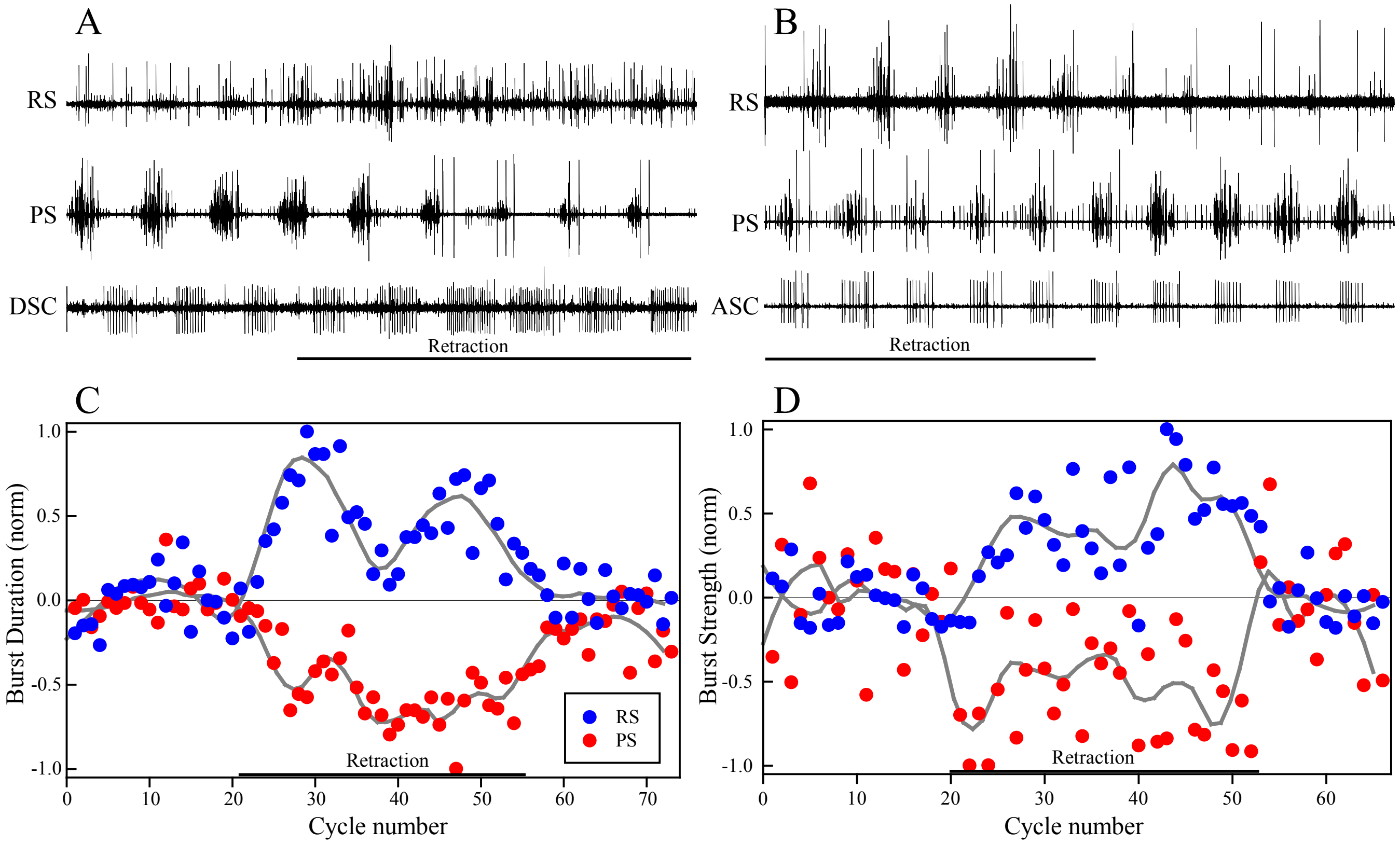

Figure 2

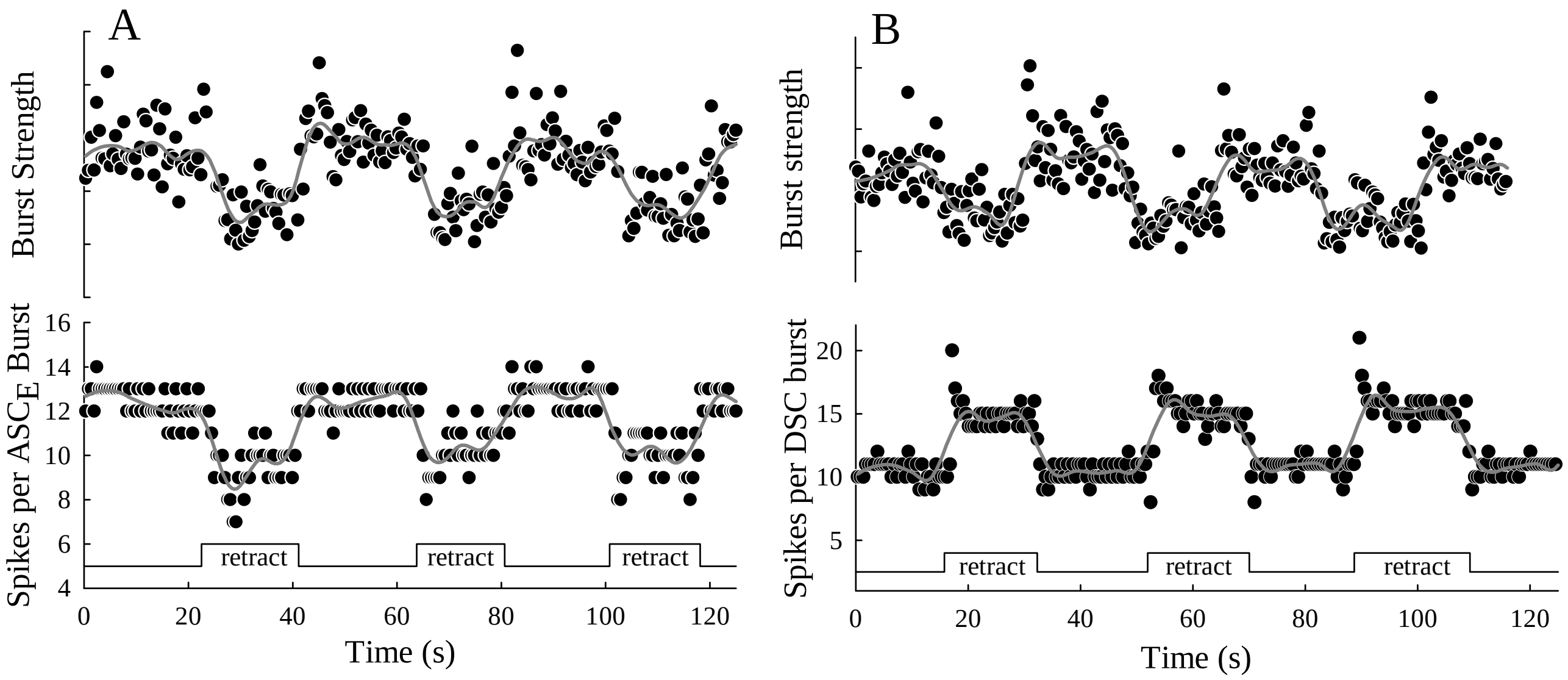

Figure 3
NSSR Figures.jnb

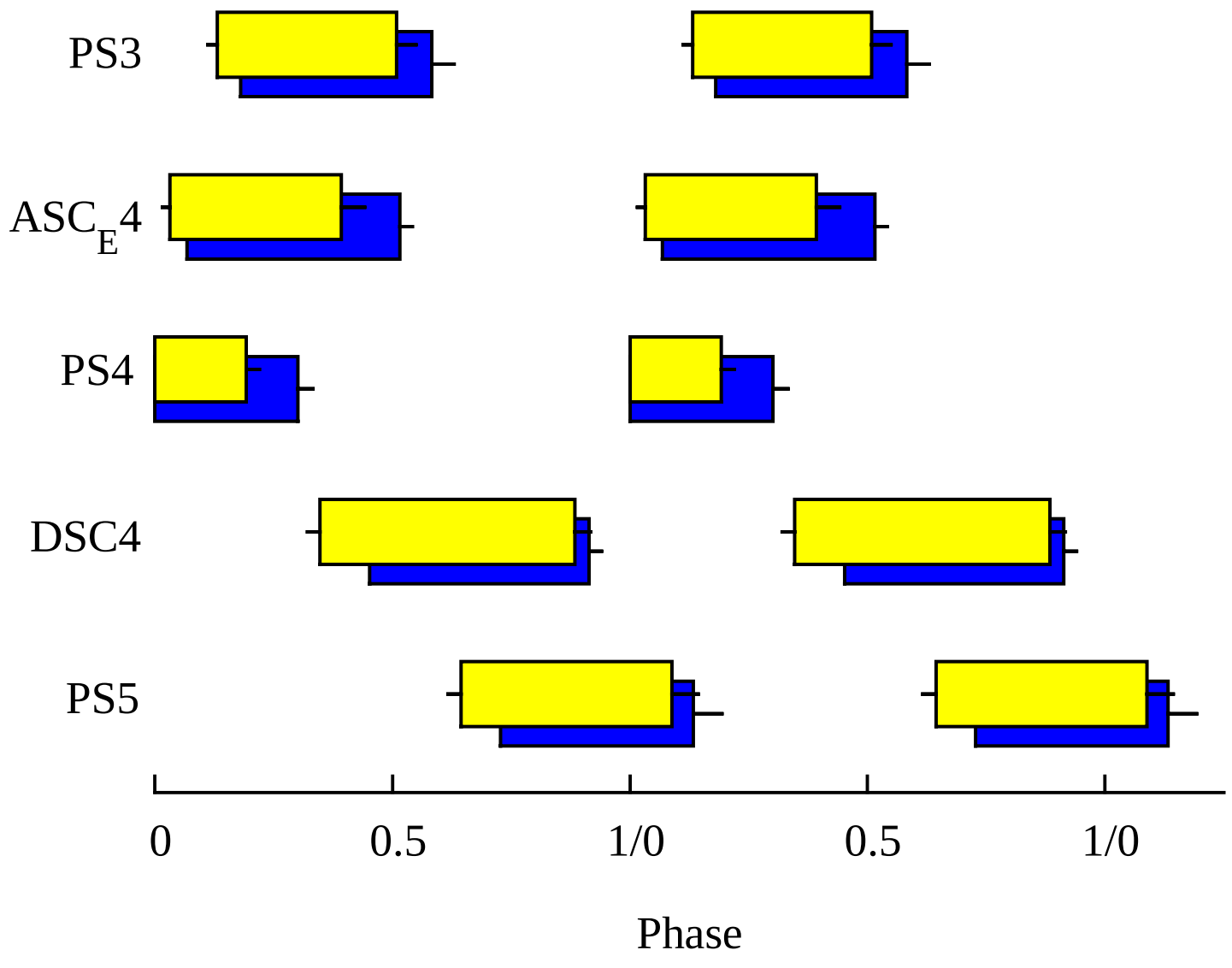

Figure 4

NSSR Figures.jnb

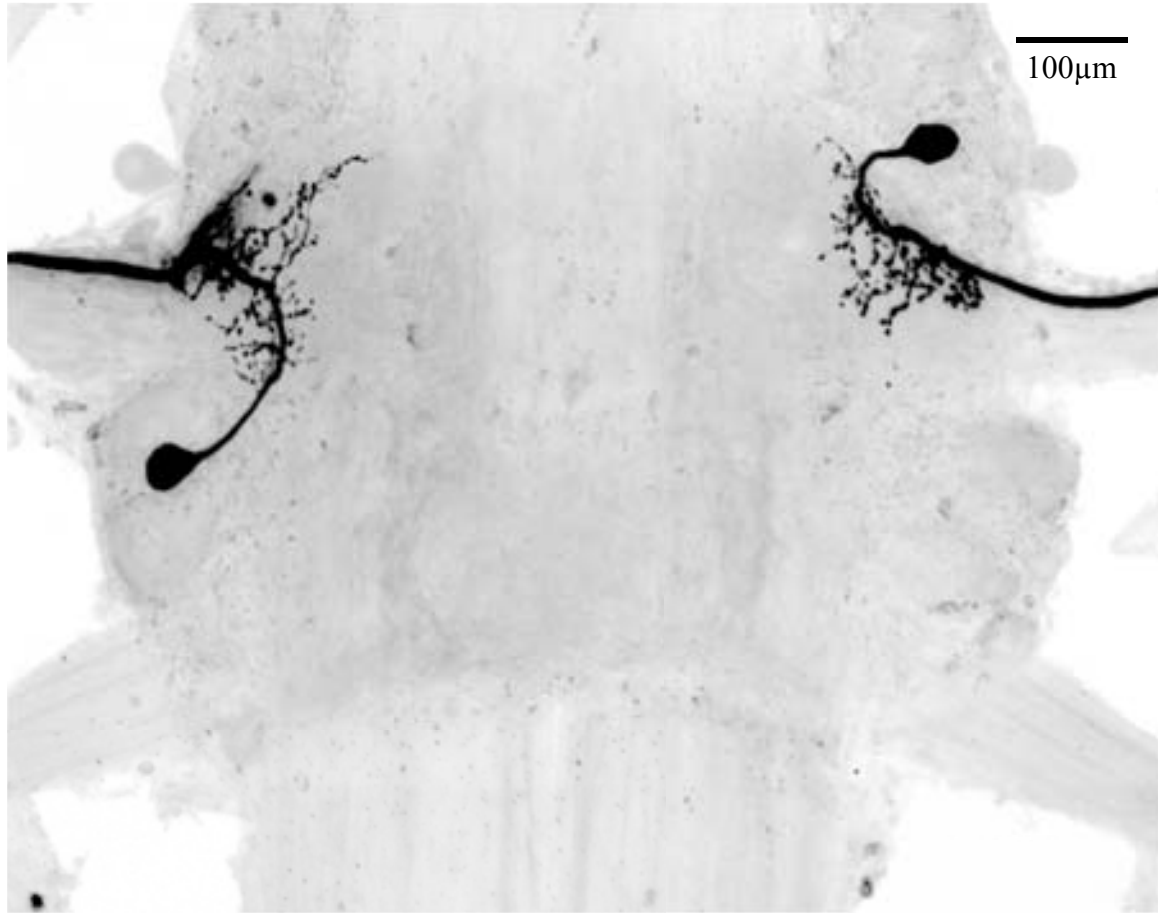

Figure 5 v3

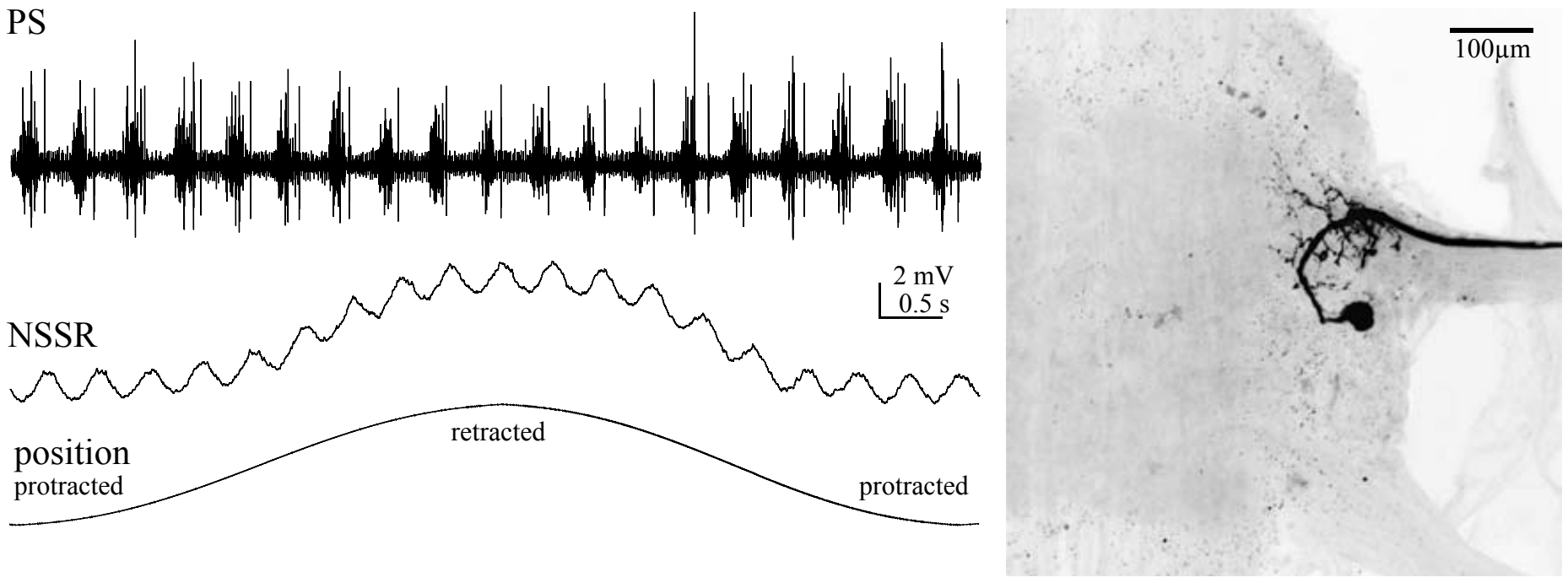

Figure 6 v2

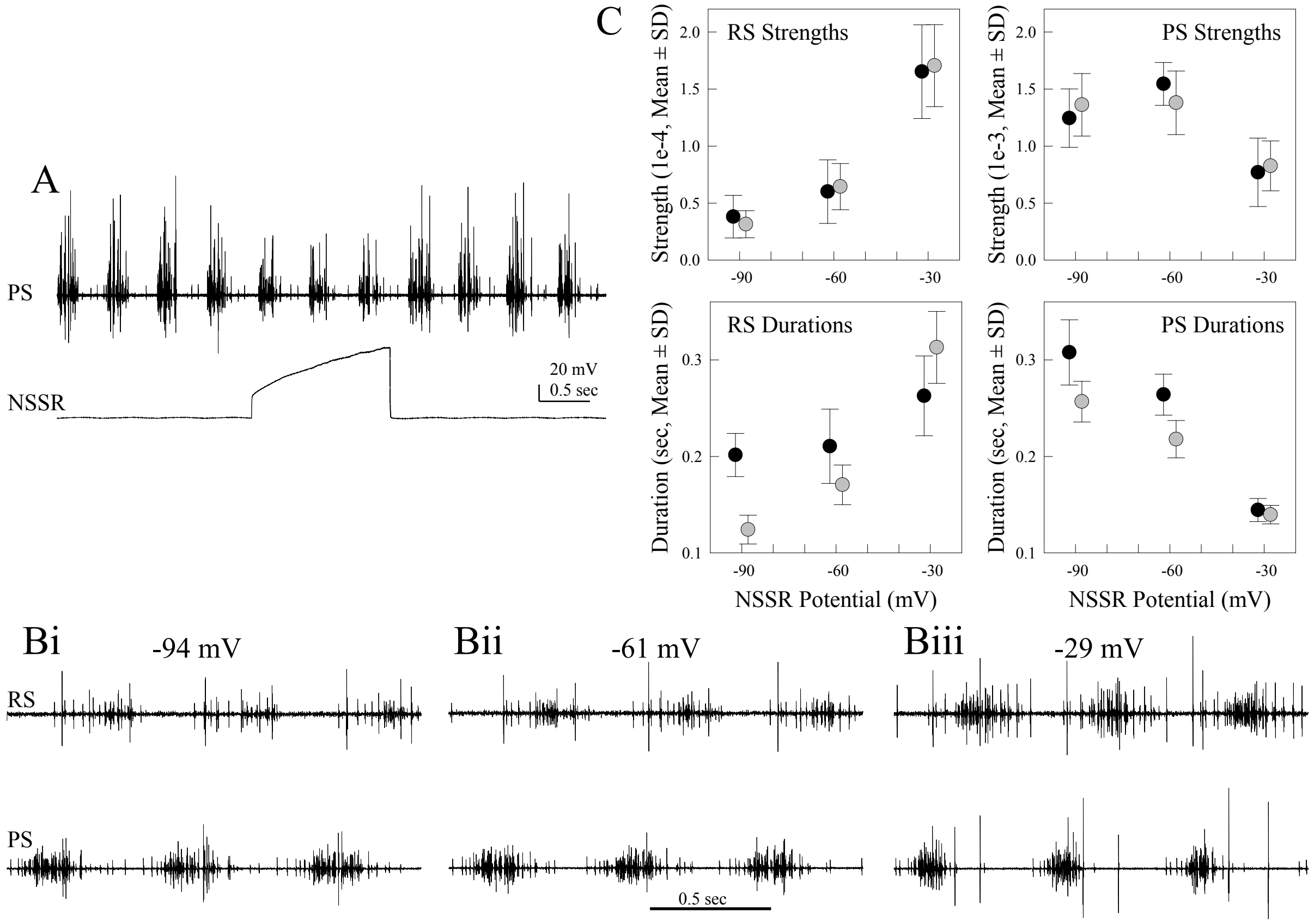

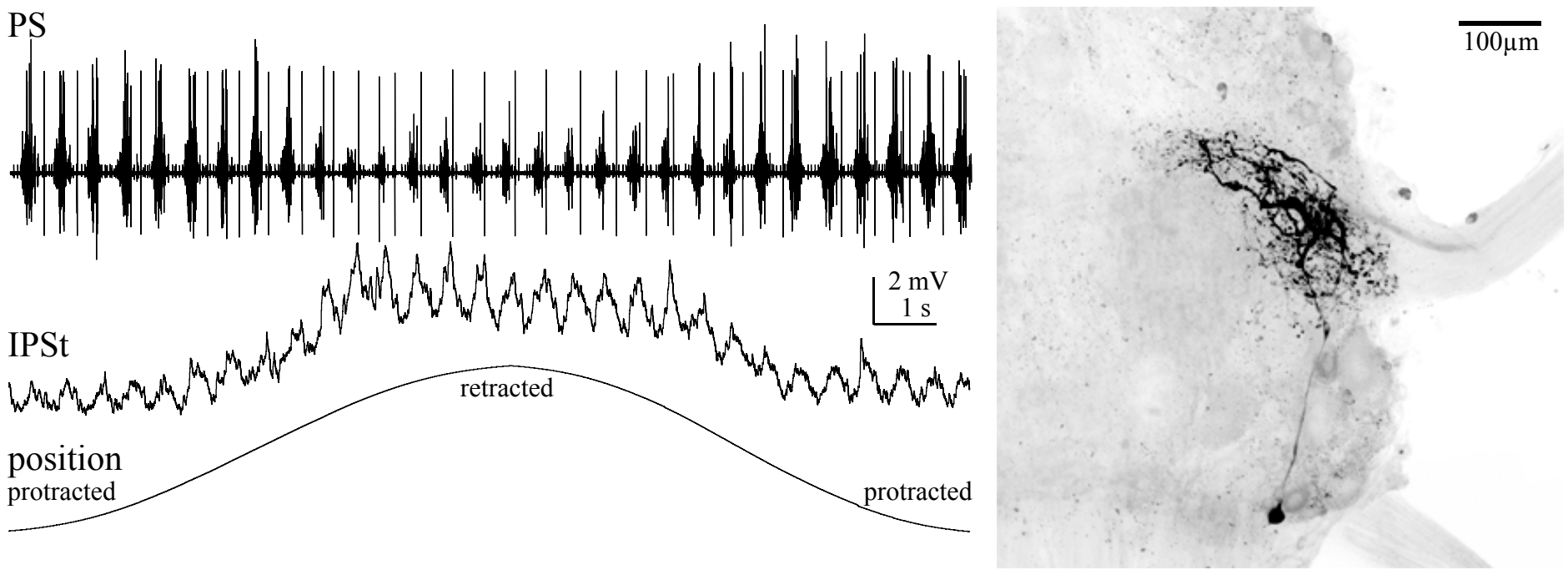

Figure 8 v2